\input epsf
\mathsurround=2pt

\def\cnt#1#2{\noindent\hangindent 4em\hangafter1\hbox to 4em{\hfil#1\quad}#2}

\newtoks\footline\footline={\ifnum\pageno>0
\hss\folio\hss\fi}
\pageno=0
~
\vskip 0in
\centerline {{\bf MINKOWSKI BESSEL MODES}\footnote*{Published in Phys. Rev. D
 {\bf38}, 514-521 (1988)}}
\bigskip

\centerline {Ulrich H. Gerlach}
\smallskip
\centerline {Department of Mathematics}
\smallskip
\centerline {Ohio State University}
\smallskip
\centerline {Columbus, Ohio  43210}
\vfill
\bigskip
\centerline {\bf ABSTRACT}
\medskip

{\narrower {\narrower \noindent
The global Minkowski Bessel (M-B) modes, whose explicit form allows the 
identification and description of the condensed vacuum state resulting from 
the operation of a pair of accelerated refrigerators, are introduced.   They 
span the representation space of the unitary representation of the Poincare 
group on 2-D Lorentz space-time.   Their three essential properties are:  
(1)  they are unitarily related to the familiar Minkowski plane waves; 
(2)  they form a unitary representation of the translation group on two 
dimensional Minkowski spacetime.  (3)  they are eigenfunctions of Lorentz 
boosts around a given reference event.  In addition the global Minkowski 
Mellin modes are introduced.  They are the singular limit of the M-B modes.  
This limit corresponds to the zero transverse momentum solutions to the zero 
rest mass wave equation.\smallskip }}

\narrower {\narrower {
\noindent
Also introduced are the four Rindler coordinate representatives of
each global mode .  Their normalization and density of states are
exhibited in a (semi-infinite) accelerated frame with a finite bottom.
In addition we exhibit the asymptotic limit as this bottom approaches
the event horizon and thereby show how a mode sum approaches a mode
integral as the frame becomes bottomless.\bigskip
}}

\centerline {\bf I.  MOTIVATION AND SUMMARY.}
\medskip

There are reasons to believe that the quantum mechanics of a
relativistic system with infinitely many degrees of freedom, e.g. the
Klein-Gordon wave field, manifests itself in a qualitatively different
way relative to a linearly uniformly accelerated frame than to
inertial frames.  Consider different inertial frames.  They are all
equivalent.  This is expressed by the fact that the ground state of a
relativistic wave field is the same relative to these inertial frames.
Thus all inertial refrigerators produce the same quantum state, the
familiar Minkowski vacuum.  By contrast a pair of refrigerators
accelerating linearly into opposite directions produce a different
quantum state.  It can perhaps best be described as a "condensed
vacuum state"$^{1,2}$.  The peculiar feature of such a state is that
even though it manifests itself in each coaccelerating frame as a no
particle state$^{3,4}$ i.e. as a vacuum, in an inertial frame it
manifests itself as liquid light in the form of a superfluid$^{1,2}$.

\medskip

What is the most direct way of identifying such a quantum state?  One
certainly could use the quantized Minkowski plane wave modes.  But
this use lacks directness.  A superior way, it turns out, is to
quantize the global Minkowski Bessel (M-B) modes.  Their existence has
in part already been known for some time$^{4,5}$, but their simple
global construction and properties as well as their usefulness as a
working tool do not yet seem familiar to theoretical physics.  The
purpose of this note is to remedy this gap.

\medskip

Minkowski Bessel modes are the global extensions of Sommerfeld's
cylinder waves to \hfil\break Minkowski space-time.  These modes allow
one to relate at a glance (a) the wave field dynamics (e.g. emission
and absorption) and its quantum properties (e.g of the ground state)
in an accelerated coordinate reference frame to (b) those in an
inertial frame.

\bigskip

Linearly uniformly accelerated observers produce worldlines in
Minkowski space-time which in Euclidean space would correspond to
circles.  This correspondence extends not only to coordinate systems
(i.e. Rindler coordinates$^7$, a type of Fermi-Walker transport$^8$
induced coordinate system, correspond to polar coordinates) but also
to the wave equation and its solutions.  Thus, corresponding to the
Klein-Gordon $(K-G)$ wave equation one has the Helmolz equation.  An
inquiry into the $K-G$ wave field (solutions) relative to an
accelerated frame demands therefore that one exhibit that which in
Euclidean space corresponds to Sommerfeld's construction of cylinder
waves from plane waves.  The Minkowski space-time analogue
corresponding to Sommerfeld's cylinder harmonics are the Minkowski
Bessel modes.  This correspondence prevails in regards to all major
properties of these modes except one: space-time has a causal
structure characterized by observer induced future and past even
horizons which partition space-time into four coordinate charts.  See
Figure 1.  

\epsffile[65 400 300 700]{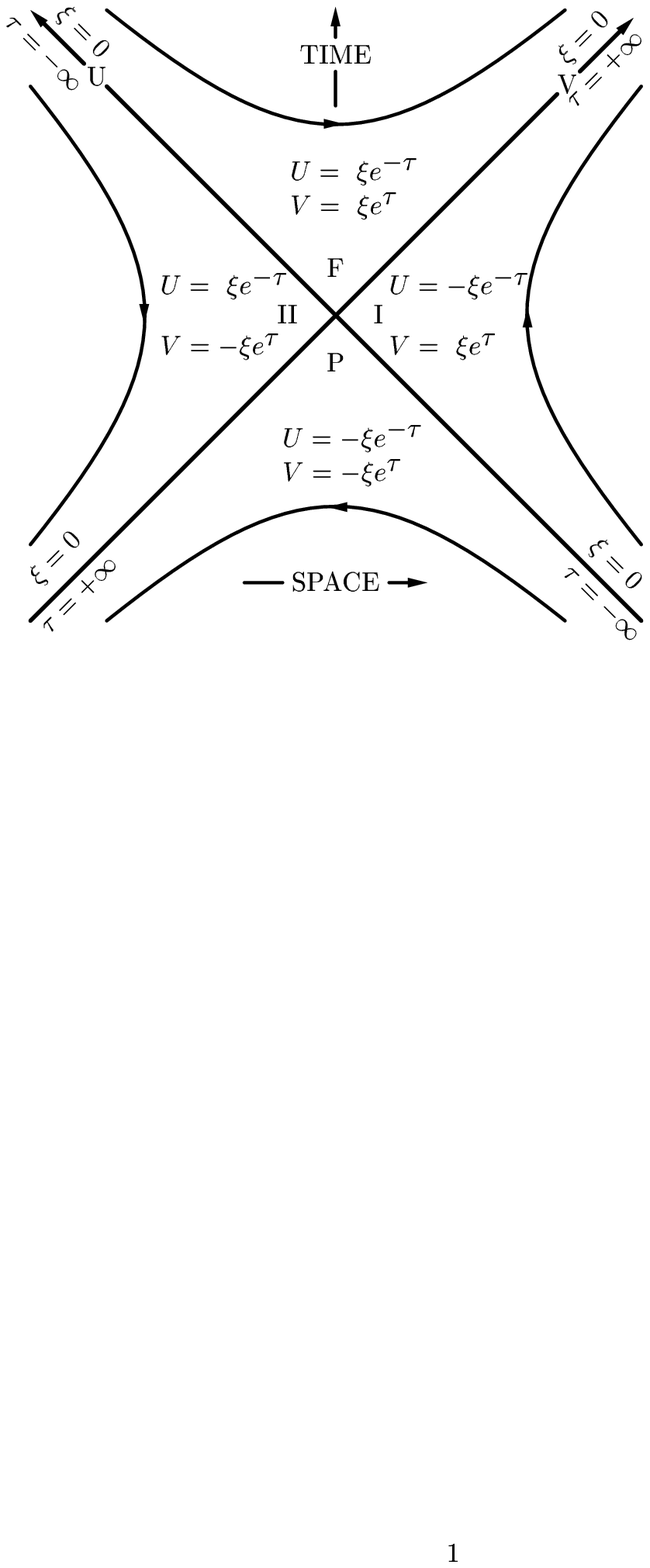}

\medskip
\centerline{\bf FIGURE 1}
\medskip
{\narrower {\narrower\smallskip\noindent
The four cooordinate neighborhoods (="Rindler" sectors) induced by the
world line of an accelerated observer.  The asymptotes $U\equiv t-x=0$
and $V\equiv t+x=0$ divide Minkowski space-times into two causally
disjoint accelerated frames I and II.  A relativistic wave field is
thereby partitioned into a pair of mutually exclusive and jointly
exhaustive subsystems.\bigskip
}}

\bigskip
Euclidean space has no such structure.  The presence of
these event horizons has a dramatic effect on the Minkowski Bessel
modes.  Because these horizons divide space-time into the four Rindler
coordinate charts, a Minkowski Bessel mode has four $\underline {\rm
coordinate\,\,representatives}$, the four Rindler modes for each of
the four Rindler sectors I, II, F and P.  The Minkowski Bessel modes
together with each of their four coordinate representatives are
pictured in Figures 2a and b.

\bigskip
\bigskip 
\eject
\noindent {\bf II.  GLOBAL PROPERTIES.}
\medskip

A Minkowski Bessel mode is a linear super-position of those plane wave modes,
$$\eqalignno {{1 \over {\sqrt {2\pi}}}\exp \lbrack \mp i(\omega_kt-k_xx)\rbrack 
\,&=\,{1 \over {\sqrt {2\pi}}}\,\exp\lbrack \mp i {k \over {2}}(Ue^\theta + 
Ve^{-\theta})\rbrack\equiv\,P^{\pm}_\theta (kU,kV)\,,&(2.1)\cr}$$
\medskip

\noindent which are on the same positive (upper sign) or negative (lower sign) 
"mass shell" given by
$$\eqalignno {\omega_k \,&=\,k \cosh \theta \,;
k_x\,=\,k \sinh \theta \,,-\infty < \theta < \infty ;\cr
k \,&=\,\vert k\vert = \sqrt {k^2_y+k^2_z+m^2}\,.&(2.2)\cr}$$
\medskip

\noindent The new coordinates 
$$U\,=\,t\,-\,x\,,\,\,V\,=\,t\,+\,x$$
\medskip

\noindent are the retarded and advanced times ("null coordinates") repectively.

A Minkowski Bessel mode in two dimensions is given by
$$\eqalignno {B^{\pm}_\omega (kU,kV)\,&=\,{1 \over {2\pi}}\int^\infty_{-\infty}
\,\exp\lbrack \mp i (\omega_k t - k_xx)\rbrack e^{-i\omega \theta }d\theta \cr
&=\,{1 \over {2\pi}}\int^\infty_{-\infty}\exp\lbrack \mp ik(Ue^\theta + 
Ve^{-\theta})/2\rbrack e^{-i\omega \theta}d\theta\cr
&= {1 \over {\sqrt {2\pi}}}\int^\infty_{-\infty}P^{\pm}_\theta (kU,kV)
e^{-i\omega \theta}d\theta &(2.3)\cr}$$
\medskip

\noindent This is the defining property .  Thus we have
\medskip

\noindent {\it Property 1}  
\medskip

The global Minkowski Bessel modes $B^{\pm}_\omega (kU,kV)$ are by means of the unitary 
transformation  $(1/2\pi)^{1/2}e^{-i\omega \theta}$ related to the plane wave 
modes $P^{\pm}_\theta (kU,kV)$.
\medskip

\noindent The upper (resp. lower) sign refers to positive (resp. negative) 
Minkowski frequency modes.  The two dimensional Klein-Gordon inner product, 
$$\eqalignno {(\psi ,\phi )\equiv i\,\int^\infty_{-\infty}\lbrack 
\psi^\ast\,\partial_t\phi\,&-\,\phi \partial_t\,\phi^\ast\rbrack dx
\equiv i\int^\infty_{-\infty}\psi^\ast \buildrel \leftrightarrow \over \partial_t 
\phi dx &(2.4)\cr}$$
\medskip

\noindent can be used to verify the sign of the Minkowski frequency of a mode.  
The inner product of two plane wave modes 
$P^{\pm}_{\theta }$ and $P^{\pm}_{\theta '}$ is
$$\eqalignno {(P^\pm_\theta ,P^\pm_{\theta '})\,&=\,\pm\,2\delta 
(\theta -\theta ')\,.&(2.5)\cr}$$
\medskip

\noindent One readily sees the well-known fact that a unitary transformation 
such as $(1/2\pi)^{1/2}e^{-i\omega \theta}$ \hfil\break preserves the $K-G$ 
inner product.  This can be verified by inserting Eq. (2.3) into Eq. (2.4) one obtains 
$$\eqalignno {(B^\pm_\omega,\,B^\pm_{\omega '})\,&=\,\pm 2\delta (\omega -\,
\omega ')&(2.6)\cr}$$
\medskip

From the point of view of quantum theory the upper plus (resp. lower minus) 
sign of the related modes $P^\pm_\theta$ and $B^\pm_\omega$ refer to a field 
whose quanta are absorbed (resp. emitted).  The absorption-emission distinction 
is the same for $P^\pm_\theta$ and $B^\pm_\omega$.  Consequently the 
quantization of the Klein-Gordon field in 
terms of the set of global Minkowski plane wave modes $P^\pm_\theta$ is 
equivalent to that in terms of the global Minkowski Bessel modes $B^\pm_\omega$.
\bigskip

\noindent {\it Property 2}
\medskip

The Minkowski Bessel modes form a (reducible) unitary representation of the translation 
group acting on the two dimensional Lorentz space-time:

$$\eqalignno {B^\pm_{\omega - \overline {\omega}}\,(k(U+U_o),k(V+V_o))\,
&=\,\int^\infty_{-\infty}B^\pm_{\omega -\omega '}(kV,kU)~B^\pm_{\omega '-
\overline {\omega}}(kU_o,kV_o)~d\omega '\,.&(2.7)\cr}$$
\medskip

\noindent This can be readily verifies by using Eq. (2.3).
\medskip

\noindent Thus
$$B^\pm_{\omega -\overline {\omega}}(kU_o, kV_o)$$
\medskip

\noindent is the unitary kernel for the space-time translation $(U_o, V_o)$.  By 
contrast the kernel for the plane wave representation is diagonal and is given by
$$P^\pm_\theta (kU,kV)\delta (\theta - \overline {\theta })\,.$$
\medskip

\noindent It satisfies an addition law analogous to Eq. (2.7).
\medskip

The plane wave modes evidently constitute irreducible representations, but the
set of Bessel modes constitutes a reducible representation of the translation 
group.
\bigskip

\noindent {\bf III.  TWO IRREDUCIBLE UNITARY REPRESENTATIONS.}
\medskip

If the group is the Poincar\'e group in two space-time dimensions then the 
$M-B$ modes yield two irreducble unitary representations.  A typical group element 
can be realized by the $3\times 3$ matrix
$$g(\tau ,t,x)\,=\,\left [ \matrix {\cosh \tau&\sinh \tau&t\cr
                                     \sinh \tau&\cosh \tau&x\cr
                                     0&0&1\cr}
\right]\,=\,g(0,t,x)~ g(\tau ,0,0)\,.$$
\medskip

\noindent The two unitary representation kernals are
$$T^{\pm}_{\omega \omega '}(\tau ,t,x)\,=\,B^{\pm}_{\omega - \omega '}
(kU,kV)e^{-i\omega '\tau}$$
\medskip

\noindent where $U\,=\,t-x$ and $V\,=\,t+x$.  The group element $g(\tau ,t,x)$ 
is the product of a pure boost
$$g(\tau ,0,0)\,=\,\left[
\matrix {\cosh \tau&\sinh \tau &0\cr
         \sinh \tau &\cosh \tau &0\cr
         0&0&1\cr}\right]$$
\medskip

\noindent  and a pure translation
$$g(0,t,x)\,=\,\left[ \matrix {1&0&t\cr
                               0&1&x\cr
                               0&0&0\cr}\right]$$
\medskip

\noindent by the amount $t\,=\,(U+V)/2, x\,=\,(U-V)/2$.  The unitary 
representation kernels corresponding to the pure boost $g(\tau ,0,0)$ are 
$$T^{\pm}_{\omega \omega '}(\tau ,0,0)\,=\,\delta (\omega - \omega ')~
e^{-i\omega \tau}\,.$$
\medskip

\noindent The kernels for the pure translation $g(0,t,x)$ are
$$T^{\pm}_{\omega ' \omega ''}(0,t,x)\,=\,B^{\pm}_{\omega ' - \omega ''}(kU,kV)$$
\medskip

An arbitrary element of the Poincre group in 2-D can always be decomposed into a 
product to a boost, a traslation, and a boost.
\medskip

In other words
$$\eqalign {&\left[ 
\matrix {\cosh (\tau-\sigma )&\sinh (\tau - \sigma)&t \cosh \tau+x\sinh \tau\cr
         \sinh (\tau - \sigma )&\cosh(\tau - \sigma )&t\sinh \tau +t\cosh \tau\cr
         0&0&1\cr}\right]\,=\cr
\cr
&=\,g(\tau ,0,0)~g(0,t,x)~g(-\sigma ,0,0)\,.\cr}$$
\medskip

\noindent The unitary representation kernels corresponding to this generic group 
element are
$$\eqalign {&\int^\infty_{-\infty} d\omega '\,\int^\infty_{-\infty}d\omega ''~\,
T^{\pm}_{\omega \omega '}(\tau, 0,0)\,~T^{\pm}_{\omega ' \omega ''}(0,t,x)~\,
T^{\pm}_{\omega '' \overline {\omega }}(-\sigma ,0,0)\,=\cr
\cr
&=\,e^{-i\omega \tau }B^{\pm}_{\omega - \overline {\omega}}(kU,kV)e^{i\overline 
{\omega }\sigma}\cr}$$
\medskip

\noindent If the two Lorentz boosts are equal and opposite, i.e. $\tau\,=\,
\sigma$, then
$$g(\tau ,0,0)~g(0,t,x)~g(-\tau ,0,0)\,=\,\left[ 
\matrix {1&0&t \cosh \tau +x\sinh \tau \cr
         0&1&t \sinh \tau +x\cosh \tau \cr
         0&0&1\cr}\right]$$
\medskip

\noindent and the corresponding unitary kernels are
$$B^{\pm}_{\omega - \overline {\omega}}(kU,kV)e^{-i(\omega - \overline 
{\omega})\tau}\,=\,B^{\pm}_{\omega - \overline {\omega}}(kUe^{-\tau},
kVe^{\tau})\,.$$
\medskip

\noindent This agrees with our physical expectations which demand that these 
kernels refer to a  translation relative to a frame Lorentz rotated (=boosted) 
by the amount $\tau$.
\medskip

There are two distinct representation spaces.  They are spanned by
$B^+_\omega (kU,kV)$ and \hfil\break $B^-_\omega (kU,kV)$ with $-\infty <\omega <\infty$. 
The two unitary representations $T^+_{\omega \overline {\omega }}$ and 
$T^-_{\omega \overline {\omega}}$ of the Poincare group act respectively 
on these two representation spaces.  The $+$ sign refers to the space of 
positive Minkowski frequencty modes, and the $-$ sign to the space of negative 
frequency modes.  One or the other set of modes plays the same role on the 
Lorentz plane in relation to the Poincare group that the familiar spherical 
harmonics play on the unit two sphere in relation to the rotation group.
\vfill\eject

\epsffile[100 100 300 650]{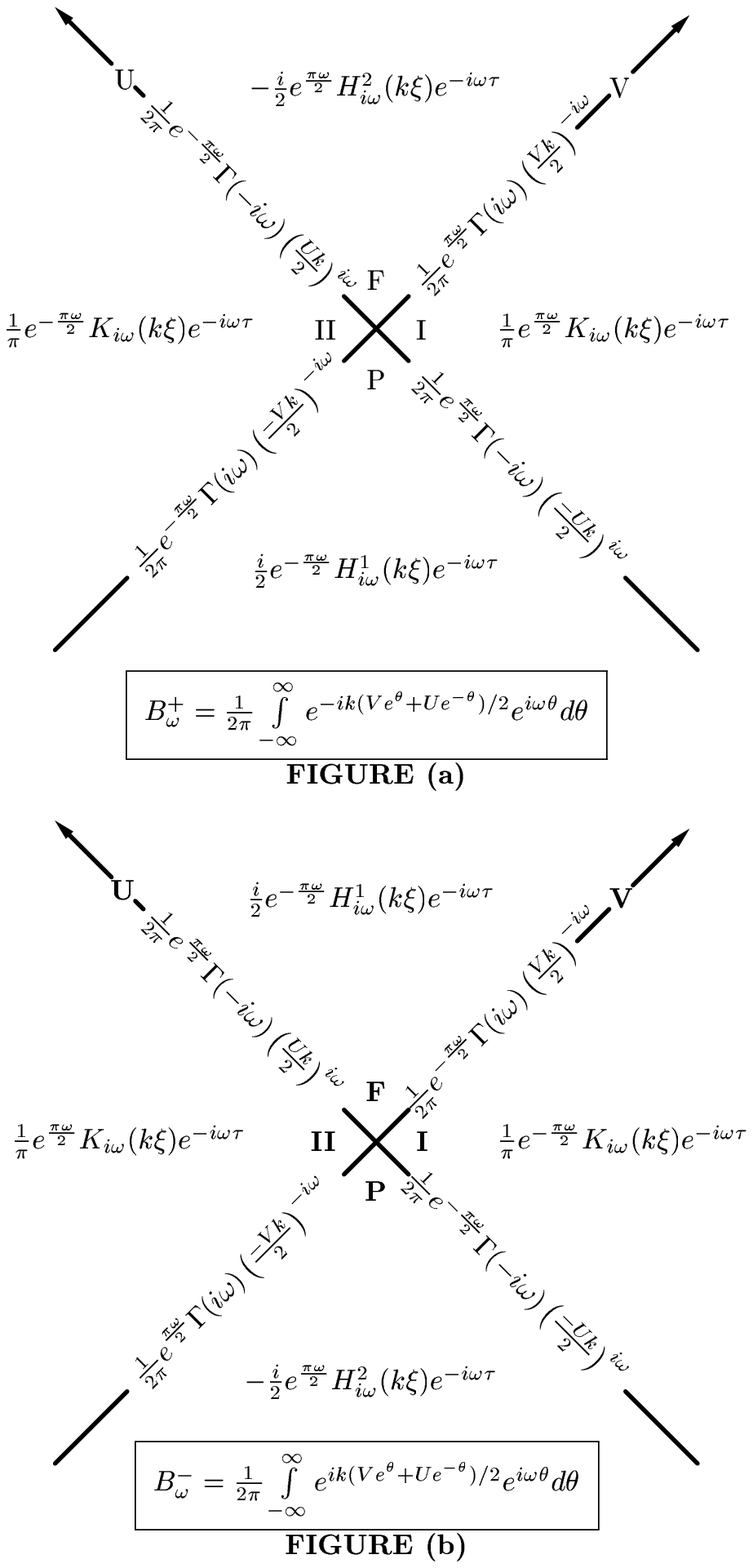}

\centerline{\bf FIGURE 2A and 2B}

\narrower{\narrower{
Global Minkowski Bessel (M-B) modes $B^{\pm}_\omega (kU,kV)$ and their
coordinate representatives.  A global M-B mode is related to its
representatives in the respective Rindler charts $I, II, F$, and $P$ by
Eqs. (4.2) - (4.3). On the past and future event horizons, $UV=0$, these 
modes have the indicated form, which is obtaied by evaluating 
the integrals in Eqs. (6.2a) and (6.2b)\bigskip
}} 

\noindent {\bf IV.  LOCAL PROPERTIES}
\medskip

A linearly uniformly accelerated observer, whose worldline is
$$\eqalign {t\,&=\,\xi  \sinh\tau ; ~~U\,=\,-\,\xi\,e^{-\tau}\cr
            x&=\,\xi \cosh\tau ;~~ V\,=\,\xi\,e^{\tau}\cr
            y\,&=\,y_0\cr
            z\,&=z_0\,,\cr}$$
\medskip

\noindent induces a division of two-dimensional Minkowski space-time into four
coordinate charts:  I, II, F and P.  See Fig. 1.  Each chart is coordinatized
by the local Rindler coordinates
$$\xi\,>\,0 ~{\rm and }~-\infty <\tau<\infty$$
\medskip

$$\eqalignno {I&:\left \{ \matrix 
{t\,&=\,&\xi \sinh\,\tau; x\,&=\,&\xi \cosh \tau\cr
U\,&=\,&-\xi e^{-\tau};\,V\,&=\,&\xi e^\tau \cr}\right.
~~~~~({\rm "right\,\,hand\,\,Rindler\,\,sector"})
&(4.1a)\cr
\cr
II&:\left \{ \matrix  
{t\,&=\,&-\xi \sinh\,\tau ; x\,&=\,&-\xi \cosh \tau\cr
U\,&=\,&\xi e^{-\tau};\,V\,&=\,&-\xi e^\tau \cr}\right.
({\rm "left\,\,hand\,\,Rindler\,\,sector"})
&(4.1b)\cr
\cr
F&: \left \{ \matrix 
{t\,&=\,&\xi \sinh\,\tau; x\,&=\,&\xi \cosh \tau\cr
U\,&=\,&\xi e^{-\tau};\,V\,&=\,&\xi e^\tau \cr}\right.
~~~~~({\rm "Future"})
&(4.1c)\cr
\cr
P&:\left \{ \matrix 
{t\,&=\,&-\xi \sinh\,\tau; x\,&=\,&-\xi \cosh \tau \cr
U\,&=\,&-\xi e^{-\tau};\,V\,&=\,&-\xi e^\tau \cr}\right.
({\rm "Past"})
&(4.1d)\cr}$$
\medskip
               
A pair of accelerated observers confine themselves to sector I and sector II
respectively.  Relative to either coordinates the metric has the form
$$ds^2\,=\,-dUdV\,+\,dy^2\,+\,dz^2\,=\,-\xi^2d\tau^2\,+\,d\xi^2\,+\,dy^2\,+\,dz^2\,.$$
\medskip

\noindent For sector F and P the metric has the form
$$ds^2\,=\,-dUdV\,+\,dy^2\,+\,dz^2\,=\,\xi^2\,d\tau^2\,-\,d\xi^2\,+\,dy^2\,+\,
dz^2\,.$$
\medskip

A global Minkoski Bessel mode, Eq. (2.3), can easily be evaluated in the 
accelered coordinate frame  sector I.  One obtains the {\it coordinate 
representative} for sector I:
$$\eqalignno {B^\pm_\omega (kU,kV)\mid_I\,&=\,(1/2\pi)\,\int^\infty_{-\infty}\,
e^{\pm ik\xi \sinh(\theta -\tau)}\,e^{-i\omega \theta}\,d\theta\,=\,{1 \over 
{\pi}}\,e^{\pm \pi \omega /2}K_{i\omega} (k\xi )^{e^{-i\omega \tau}}&(4.2a)\cr}$$
\medskip

Simlilarly for the left hand sector II, the future quadrant $F$, and the past 
quadrant $P$ one has the respective coordinate representatives
$$\eqalignno {B^\pm_\omega (kU,kV)\mid_{II}\,&=\,(1/2\pi)\,\int^\infty_{-\infty}\,
e^{\mp ik\xi \sinh(\theta -\tau)}\,e^{-i\omega \theta}\,d\theta\,=\,{1 \over 
{\pi}}\,e^{\mp \pi \omega /2}K_{i\omega} (k\xi )^{e^{-i\omega \tau}},&(4.2b)\cr}$$
\medskip

$$\eqalignno {B^{\pm}_\omega (kU,kV)\mid_F\,&=\,(1/2\pi)\,\int^\infty_{-\infty}\,
e^{\mp ik\xi \cosh(\theta -\tau)}\,e^{-i\omega \theta}\,d\theta\,=\,\mp\,{i \over {2}}
e^{ \pm\pi \omega /2} H^{2 \atop 1}_{i\omega }(k\xi )e^{-i\omega \tau},&(4.3a)\cr}$$
\medskip

\noindent and

$$\eqalignno {B^{\pm}_\omega (kU,kV)\mid_P\,&=\,(1/2\pi)\,\int^\infty_{-\infty}\,
e^{\mp ik\xi \cosh(\theta -\tau)}\,e^{-i\omega \theta}\,d\theta\,=\,\pm {i \over {2}} 
e^{\mp \pi \omega /2} H^{1 \atop 2}_{i\omega }(k\xi )e^{-i\omega \tau}\,.&(4.3b)\cr}$$
\medskip

\noindent These coordinate representatives are Sommerfeld's cyclinder waves 
generalized from Euclidean space coordinatized by polar coordinates to Minkowski 
space-time coordinatized by Rindler coordinates (See Figure 2).  These waves are 
expressed in terms of 
$$ K_{i\omega }(k\xi)\,=\,K_{-i\omega} (k\xi)\,,$$
\medskip

\noindent the Bessel functions of imaginary argument $ik\xi$ 
("MacDonald functions") and imaginary 
order $i\omega$.  In F and P the waves are expressed in terms of
$$e^{\pi \omega /2}H^{1 \atop 2}_{i\omega}\,=\,e^{-\pi \omega /2}\,H^
{1 \atop 2}_{-i\omega}(k\xi )\,,$$
\medskip

\noindent multiples of the two kinds of Hankel functions.  All four coordinate representatives are multiples of $e^{-i\omega \tau}$.  
Consequently, each one is an eigenfunction of the Lorentz boost around the 
reference event $t\,=\,x\,=\,0$.  But they all have the same eigenvalue.  Thus 
one has
\bigskip
\noindent {\it Property 3}
\medskip
$$\eqalignno {{\partial \over {\partial \tau}}B^\pm_{i\omega}\,&=\,-i\omega 
B^\pm_{i\omega}&(4.4)\cr}$$
\medskip

\noindent in all four coordinate neighborhoods, i.e. a globally defined Minkowski 
Bessel mode (M-B)is an eigenfunction of the Lorentz boost operation.
\medskip

One can see that a Lorentz boost in an inertial frame is simply a time $(\tau )$
translation in the accelerated frame.
\vfill\eject

\noindent {\bf V.  ORTHONORMALIZATION.}
\medskip

The utility of the Klein-Gordon ("Wronskian") orthonormality, Eq. (2.6) 
$$\eqalignno {\pm 2\delta (\omega - \omega ')\,&=\,i
\int\limits_{\scriptstyle space-like\atop \scriptstyle hypersurface} 
(B^{\pm}_{i\omega})^\ast ~{\buildrel \leftrightarrow \over 
\partial_\mu} B^{\pm}_{i\omega '}~d^1\Sigma^\mu\,,&(5.1)\cr}$$ 
\medskip

\noindent of the $M-B$ modes extends beyond their identification as positive and 
negative Minkowski frequency modes.  One can also use it to obtain a very useful 
normalization integral in the accelerated frame. We shall do this for a frame 
without any bottom $(0<\xi <\infty )$ as well as for a frame with a bottom 
$(b\le \xi <\infty\,,\,\,b>0)$.  These results we shall use to obtain the 
asymptotic behavior of mode sums in an accelerated frame with a bottom.  We take 
the limit as the bottom aproaches the event horizon $(\xi = 0)$ and the frame 
becomes thereby bottomless.
\medskip

\noindent {\it A. Bottomless Frame $(0<\xi <\infty )$}
\medskip

The result is obtained in three steps.   
\medskip

\noindent (1)  Evaluate the hypersurface integral, Eq. (5.1) on the (1-dimensional) 
space-like hypersurface $\tau =$ constant in both Rindler sectors I and II.  Thus 
the only non-zero component of the hypersurface element
$$d^1\Sigma^\mu\,=\,g^{\mu \nu}\epsilon_{\nu \alpha}dx^\alpha\,=\,g^{\mu \nu}
\sqrt {-g}\lbrack \nu , \alpha \rbrack dx^\alpha$$
\noindent is
$$d^1\Sigma^\tau\,=\,{d\xi \over {\xi}}\,.$$
\medskip

The Wronskian integral, Eq. (5.1) can therefore be written as
$$\eqalignno {2\delta (\omega - \omega ')\,=\,i\int^0_\infty (B^+_{i\omega})^\ast
{{\buildrel \leftrightarrow \over \partial} \over {\partial \tau}}B^+_{i\omega '}
\vert_{II}~{d\xi \over {\xi}}\,&+\,i\int^\infty_0(B^+_{i\omega})^\ast {{\buildrel 
\leftrightarrow \over \partial} \over {\partial \tau}}B^+_{i\omega '}\vert_I~
{d\xi \over {\xi}}&(5.2)\cr}$$
\medskip

\noindent It is unnecessary to use $B^-_{i\omega}$ because it will give the same 
result.
\medskip

\noindent (2)  Use Property 3 and insert the coordinate representatives, Eqs. 
(4.2a) and (4.2b) into Eq. (5.2).  The result is
$$\eqalignno {\delta (\omega - \omega ')\,&=\,{\omega + \omega ' \over {\pi^2}}
\sinh \lbrack {\pi (\omega + \omega ') \over {2}}\rbrack e^{i(\omega '-\omega)
\tau}\int^\infty_0~K_{i\omega}(k\xi )K_{i\omega '}(k\xi)~{d\xi \over {\xi}}
&(5.3)\cr}$$
\medskip

Our interest lies in the integral, which  is not determined by this equation when 
$\omega + \omega '\,=\,0$.  This however is not a problem because 
$K_{i\omega}(k\xi)$ is an even function of $\omega $, i.e.
$$K_{-i\omega}(k\xi)\,=\,K_{i\omega}(k\xi)\,.$$
\medskip

\noindent Consequently
$$\eqalignno {\delta (\omega + \omega ')\,&=\,{\omega - \omega '\over {\pi^2}}
\sinh \lbrack {\pi (\omega - \omega ') \over {2}}\rbrack ~e^{-i(\omega + 
\omega ')\tau}\int^\infty_0\,K_{i\omega}(k\xi)K_{i\omega '}(k\xi){d\xi \over 
{\xi}}\,.&(5.4)\cr}$$
\medskip

\noindent (3)  Add Eqs. (5.3) and (5.4) to obtain the useful normalization 
integral 
$$\eqalignno {\int^\infty_0~K_{i\omega }(k\xi)~K_{i\omega '}(k\xi ){d\xi \over {\xi}}\,&=\,{\pi^2 \over 
{2\omega \sin h \pi \omega}}\lbrack \delta (\omega - \omega ')+\delta 
(\omega + \omega ')\rbrack \,.&(5.5)\cr}$$
\medskip

The fact that both delta functions occur on the right side is a reflection of the  
fact that the Bessel function $K_{i\omega }(k\xi)$ is an even function of 
$\omega$.
\bigskip

\noindent {\it B. Frame With a Finite Bottom} $(b\le \xi <\infty )$
\medskip
  
We consider the mode $K_{i\omega }(k\xi )$ which satisfies the differential 
equation $(k^2=k^2_y+k^2_z+m^2)$
$$\eqalignno {\lbrace \xi {d \over {d\xi }}\xi\,{d \over {d\xi }} + \omega^2 - 
k^2\xi^2\rbrace K_{i\omega}(k\xi )&=\,0&(5.6)\cr}$$
\medskip

\noindent on the domain $b\le \xi <\infty , b>0$.  It satisfies some fixed and 
given homogeneous boundary condition at $\xi = b$,
$$a_1\,K_{i\omega }(kb)+a_2 {d \over {d\xi }}K_{i\omega }(kb)\,=\,0\,.$$
\medskip
\noindent Consequently the allowed modes have discrete frequencies $\omega$.  Let 
these frequencies be
$$0< \omega_1<\omega_2< \cdots <\omega_n<\cdots \,.$$
\medskip

\noindent The Sturm Liouville nature of this eigenvalue problem guarantees that 
these modes satisfy
$$\eqalignno {\int^\infty_{kb}\,K_{i\omega_m}(x)K_{i\omega_n}(x) {dx 
\over {x}}\,
&=\,c_n(kb)\delta_{mn}&(5.7)\cr}$$
\medskip

\noindent Our objective is to obtain the normalization constant $c_n(kb)$ 
as $kb\to 0$.  Thus one can make a transition from an accelerated form with a finite 
bottom $(b>0)$ to a bottomless one $(b=0)$.  In quantum field theory or in 
condensed matter physics such a transition is called "going to the 
thermodynamic limit".  Comparing Eq. (5.7) with (5.5) one writes this 
transition as 
$$\eqalignno {c_n(kb)\delta_{mn}\to {\pi^2 \over {2\omega \sinh \pi \omega }}
\delta (\omega - \omega ')~~~\,&(as~b\to 0)&(5.8)\cr}$$
\medskip

\noindent This is a useful equation because one can now work with finite 
quantities (namely the righthand side) which in the thermodynamic limit become 
infinite (namely the l.h.s. when $\omega =\omega ')$.
\medskip

One can evaluate Eq. (5.7) exactly in terms of $K_{i\omega}$ and its 
derivative with respect to $\omega\,.^9$  But we shall use the $WKB$ approximation 
because it is more transparent.  In this approximation $(k\xi \equiv x)$
$$\eqalignno {K_{i\omega}(x)\,&=\,\sqrt {{\pi \over {2\sinh \pi \omega}}}\,
{1 \over {(\omega ^2 - x^2)^{1/4}}}\,cos(\int\,\sqrt {\omega^2-x^2}\,
{dx \over {x}}+const)&(5.9)\cr}$$
\medskip

\noindent whenever $k\xi <<\omega$.
\medskip

\noindent Consequently
$$\eqalignno {\int^\infty_{kb}\,K^2_{i\omega }(x){dx \over {x}}\,&\simeq\, 
\int^\omega_{kb}\,{\pi \over {2 \sinh \pi \omega }}\,{1 \over {\sqrt 
{\omega^2-x^2}}}\,{1 \over {2}}\lbrack 1 + cos(\cdots )\rbrack {dx \over {x}}.
&(5.10)\cr}$$
\medskip

\noindent Upon integration the term $cos(\cdots )$ averages to zero.  One 
obtains therefore
$$\eqalignno {\int^\infty_{kb}\,K^2_{i\omega }(x){dx \over {x}}\,
&\simeq\,{\pi \over {2\omega \sinh \pi \omega }}\,log {(\omega + \sqrt
{\omega^2-k^2b^2})\over {kb}},~~~kb \le \omega <\infty
&(5.11)\cr}$$
\medskip

\noindent  Using this normalization integral to compare Eq. (5.7) with (5.5) 
one obtains the desired relation for Eq. (5.8), namely
$$\eqalignno {\lim_{b\to 0}{1 \over {\pi}}\,log {(\omega + \sqrt {\omega^2-k^2b^2})\over {kb}}\delta_{mn}\,
&=\,\delta (\omega_m - \omega_n)&(5.12)\cr}$$
\medskip
\noindent {\it C. Mode Sums in an Accelerated Frame.}
\medskip

\noindent Our final objective is to establish the corresponding relation 
between a mode sum and a mode integral.  The allowed normal mode frequencies 
$\omega$ are 
determined by the "Bohr quantization" condition applied to Eq. (5.9).  One has
$$\int^\omega _{kb}\,\sqrt {\omega^2-x^2}\,{dx \over {x}}+const\,=\,n\pi$$
\medskip

\noindent The density of states is obtained by differentiation with respect to 
$\omega$ and then doing the integration, 
$$\eqalignno {{dn (\omega ) \over {d\omega }}\,&=\,{1 \over {\pi}}\,log 
{(\omega + \sqrt {\omega^2-k^2b^2})\over {kb}},~~~kb\le \omega <\infty &(5.13)\cr}$$
\medskip

\noindent It follows from Eq. (5.13) that the transition from a mode sum to a 
mode integral is established  for small $kb$ by
$$\eqalignno {\sum^\infty_{n=1}(\cdots )\buildrel kb~small \over \longrightarrow 
\int dn(\cdots )&=\, \int^\infty_{kb}d\omega {dn \over {d\omega}}(\cdots)\cr
&=\,\int^\infty_{kb}d\omega {1 \over {\pi}} log {\omega + \sqrt 
{\omega^2-k^2b^2} \over {kb}}(\cdots )&(5.14)\cr}$$
\medskip

\noindent Combining Eqs. (5.12) and (5.13) one obtains the expected result
$$\eqalignno {\sum^\infty_{n=1}\delta_{mn}(\cdots )\buildrel kb~small \over 
\longrightarrow &\int^\infty_{kb}d\omega \delta (\omega - \omega_m)(\cdots)\,.
&(5.15)\cr}$$
\medskip

\noindent Equations (5.14) and (5.15) show how mode sums for a bottomless 
$(b=0)$ accelerated frame are the asymptotic limit of corresponding sums for 
accelerated frame with a bottom $(b>0)$.
\medskip

Equations (5.12), (5.13), and (5.14) have their anologue in carteseian 
coordinates where the modes satisfy the differential equation
$$\lbrack {\partial^2 \over {\partial y^2}}+{\partial^2 \over {\partial z^2}}
+k^2_y +k^2_z\rbrack \psi = 0$$
\medskip

\noindent on the domain $- {L \over {2}}\le y,z \le {L \over {2}}$.  The 
corresponding equations are well known and are given by
$$\eqalignno {\lim\limits_{L\to \infty}({L \over {2\pi }})^2\delta_{m_yn_y}
\delta_{m_zn_z}&=\delta 
(k_y-k'_y)\delta (k_z-k'_z)&(5.12')\cr
\cr
{d(modes) \over {d k_ydk_z}}&=({L \over {2\pi}})^2&(5.13')\cr
\cr
\sum^\infty_{n_y=-\infty}\sum^\infty_{n_z =-\infty}(\cdots)\to 
\int^\infty_{-\infty}\int^\infty_{-\infty}dn_ydn_z~(\cdots )
&=\lim\limits_{L\to \infty}L^2\int^\infty_{-\infty}\int^\infty_{-\infty}
{d^2k \over {(2\pi )^2}}(\cdots )&(5.14')\cr}$$
\medskip

To achieve our final objective of relating a mode sum in a bottomless frame to 
a mode integral in a frame with a bottom one first multiplies Eqs. (5.13) 
and (5.13$'$),
$$\eqalignno {{d(all~modes)\over {(dk_y)(dk_z)(d\omega)}}\,&=\,({L \over {2\pi }})^2
{1 \over {\pi}} log{\omega +\sqrt {\omega^2 - k^2b^2} \over {kb}}&(5.16)\cr}$$
\medskip

This is the "density of all states" in a frame whose bottom is at $\xi =b$.  
Secondly one uses this density to evaluate the mode sum
$$\sum^\infty_{n_y=-\infty}\sum^\infty_{n_z=-\infty}\sum^\infty_{n=1}(\cdots)
\equiv \sum (\cdots)$$
\medskip

\noindent asymptotically for "small" $b$.  Small $b$ means
$$0<b<<g^{-1}$$
\medskip

\noindent where $g^{-1}$ is the Fermi-Walker$^8$ ("head start") distance of the 
fiducial observer whose world line is $x^2-t^2=\xi^{-2}=g^{-2}$.  As one 
readily sees from the metric
$$ds^2=-\xi^2g^2d\tau^2_{rel}+d\xi^2+dy^2+dz^2$$
\medskip

\noindent relative to the accelerated frame, this distance is at $\xi = g^{-1}$.  
There the coordinate time coincides with proper time $(\bigtriangleup s= 
\bigtriangleup 
\tau_{rel})$ and the proper acceleration is measured to be $g$.  Throughout 
this paper we have hidden this acceleration by absorbing it with the 
realtivistic "boost" time and "boost" frequency into the dimensionless 
geometrical quantities and 
$$\eqalign{\omega\,&=\,\omega_{rel}/g\cr
\cr
\tau\,&=\,\tau_{rel}g\cr}$$
\medskip

\noindent respectively.  For the purpose of exhibiting the asymptotic 
$(b<<g^{-1})$ expression of the total mode sum $\sum(\cdots )$, we shall 
temporarily reintroduce this acceleration explicitly.  Thus we have
$$\sum (\cdots )\to \int^\infty_{-\infty}\int^\infty_{-\infty}
\int^\infty_{bg}{d (all~modes) \over {(dk_y)(dk_z)(d\omega )}}dk_y dk_z d
\omega ~~~(\cdots )$$
\medskip

\noindent where the density of all moes is given by Eq. (5.16).  Introducing 
$\Omega$ by means of 
$$\omega = {\Omega k\over {g}}$$
\medskip

\noindent one obtains for $bg<<1$  
$$\eqalignno {\sum (&\cdots )\to \int^\infty_{-\infty}\int^\infty_{-\infty}
\int^\infty_{bg}({L \over {2\pi}})^2 {1 \over {\pi}} log {\Omega +\sqrt 
{\Omega^2 - b^2g^2}\over {bg}}{k \over {g}}dk_ydk_zd\Omega~~~(\cdots )&(5.17)\cr}$$
\medskip

\noindent The logarithmic factor is independent of $k_y$ and $k_z$.  
Furthermore the convergence of the \hfil\break $\int^\infty \cdots d\Omega$ 
integral implies that for $bg<<1$  this logarithmic factor becomes
$$log {\Omega +\sqrt {\Omega^2-b^2g^2} \over {bg}}\,\buildrel bg<<1 \over 
\longrightarrow\,log\,{1 \over {bg}}\,.$$
\medskip

\noindent Consequently the total mode sum is
$$\eqalignno {\sum &(\cdots ) ~~\buildrel bg<<1 \over \longrightarrow ~~L^2 log 
(1/bg)\int^\infty_{-\infty}\int^\infty_{-\infty}\int^\infty_0 {d^2k \over 
{(2\pi )^2}} {d\omega \over {\pi}} (\cdots ),&(5.18)\cr}$$
\medskip

\noindent provided $(\cdots )$ is well-behaved near $\omega = 0$.
\medskip

\noindent This is the asymptotic expression for the mode integral in an uniformly
and linearly accelerated frame.  By contrast the corresponding familiar 
expression in an inertial frame  is
$$\sum (\cdots ) ~\buildrel V~large \over \longrightarrow ~V~\int^\infty_{-\infty}
\int^\infty_{-\infty}\int^\infty_{-\infty}{d^2k \over {(2\pi )^2}}{dk_x \over 
{2\pi}}(\cdots )\,.$$
\medskip

\noindent Here $V=L^3$.  One sees that what corresponds to going to the infinite volume 
limit in an inertial frame $(V\to \infty)$, corresponds to 
$$\eqalignno {V_{R-W}\,\equiv\,L^2g^{-1}log(1/bg)&\to \infty &(5.19)\cr}$$
\medskip

\noindent in an accelerated frame.  Here $V_{R-W}$ is the Regge-Wheeler 
volume.  It's longitudinal length is based on the flat 
space-time analogue of the Regge-Wheeler ("tortoise") coordinate$^{10}$,
$\xi^\ast$,
$$\eqalignno {\xi^\ast (\xi )&= g^{-1}log \xi g.&(5.20)\cr}$$
\medskip

The Regge-Wheeler coordinate straightens out the null cone along the acceleration 
direction in an accelerated frame, be it near a black hole or in flat space 
time,
$$ds^2 = g^2\xi^2(-d\tau^2_{rel}\,+\,d\xi^{\ast 2})+dy^2 + dz^2\,.$$
\medskip

\noindent This coordinate is not proper distance.  It pushes the event horizon 
$\xi = 0$ to $\xi^\ast =- \infty$.  In terms of this coordinate $\xi = 0$ 
lies at negative spatial infinity of the accelerated frame, 
and the {\it Regge-Wheeler size} of the proper interval $\lbrack b,g^{-1}\rbrack$ is from Eq. (5.20)
$$\xi^\ast(g^{-1}) - \xi^\ast (b)= g^{-1}log(1/bg).$$
\medskip

\noindent Thus the Regge-Wheeler volume of a semi-finite accelerated box with 
bottom at $\xi = b>0$ is the product of this length with the transverse area 
$L^2$, 
$$\eqalignno {V_{R-W}\,&=\,L^2 g^{-1}log (1/bg)&(5.21)\cr}$$
\medskip

\noindent One concludes 
therefore that the "thermodynamic" limit in an accelerated frame consists of 
$$\eqalignno {\Sigma (\cdots ) ~~\buildrel V_{R-W}~large \over \longrightarrow~~
(gV_{R-W})&\int^\infty_{-\infty}\int^\infty_{-\infty}\int^\infty_0{d^2k \over 
{(2\pi )^2}}{d\omega \over {\pi}}~~(\cdots )\,.&(5.22)\cr}$$
\vfill\eject

\noindent {\bf VI.  THE DEGENERATE CASE}
\medskip

The Minkowski Bessel modes, as well as the Minkowski plane wave modes satisfy 
any one of th three equations, 
$$\eqalignno {\lbrace - {\partial^2 \over {\partial t^2}}+\,{\partial^2 \over 
{\partial x^2}}\,-\,k^2\rbrace &\psi\,=\,0\cr
\lbrace - 4 {\partial^2 \over {\partial U\partial V}}\,-\,k^2\rbrace 
&\psi\,=\,0 &(6.1)\cr
\lbrace \mp {1 \over  {\xi^2}}{\partial ^2 \over {\partial \tau^2}}\,\pm\,
{1 \over {\xi}}{\partial \over {\partial \xi}}\xi {\partial \over 
{\partial \xi}}\,-\,k^2\rbrace &\psi\,=\,0 \cr}$$
\medskip

\noindent (upper sign for $I\cup II$, lower sign for $F\cup P$) depending on 
which coordinates one uses.  These equations are the result of solving the 
Klein-Gordon equation so that
$$k^2\,=\,k^2_y\,+\,k^2_z\,+\,{m^2c^2 \over {h^2}}$$
\medskip

\noindent where the individual terms have the usual meaning.
\medskip

In this paper we have exhibited the set of Minkowski Bessel modes for the 
non-degenerate case $k^2>0$.  The degenerate case is $k^2\to 0$.  This is a set 
of measure zero and it is in a class by itself.  It corresponds to plane waves 
of a massless field travelling strictly along the $x$-direction.  There are several ways of 
obtaining the solution corresponding to this singular limit.  One of the most 
direct ways is to simply consider
$$\eqalignno {M^{r\pm}_\omega (\kappa U)\,&\equiv\,B^{\pm}_\omega (2\kappa U,0)\cr
&=\,{1 \over {2\pi }}\int^\infty_{-\infty}exp(\mp i \kappa Ue^\theta)e^{-i\omega 
\theta}d\theta&(6.2a)\cr}$$
\medskip

\noindent We shall call these modes the "retarded" (superscript "$r$") Minkowski 
Mellin (M-M) modes because they are (with $s\,=\,e^\theta$) a Mellin synthesis 
of the plane waves $exp\lbrack \mp\,i(t-x){\cal {H}}e^\theta \rbrack$.  
The "advanced" (superscript "$a$") Minkowski Mellin modes are 
$$\eqalignno {M^{a\pm}_\omega (\kappa V)\,&\equiv\,B^{\pm}_\omega (0, 2\kappa V)\cr
&=\,{1 \over {2\pi}}\int^\infty_{-\infty}exp (\mp i\kappa Ve^{-\theta})e^{-i\omega 
\theta} d\theta &(6.2b)\cr}$$
\medskip

\noindent and they are composed of positive (upper sign) or negative (lower sign) 
Minkowski frequency plane waves travelling towards negative $x$.  These Minkowski 
Mellin modes satisfy Eqs. 6.1 with $k^2\,=\,0$.  The constant $k >0$ appearing in 
Eqs. (6.2) is arbirary and has been introduced for dimensional reasons.  These 
$M-M$have first been exhibited by R. Hughes$^{11}$.  Note that the advanced 
modes considered as functions of their argument are related to the 
retarded ones 
by
$$M^{r\pm}_\omega\,=\,M^{a \pm}_{-\omega}$$
\medskip

The most important aspect of these $M-M$ modes is that they lack no 
property which the Minkowski Bessel functions have, namely
\medskip

\noindent (1)  They are unitarily related to the plane waves and hence are 
globally defined.
\medskip

\noindent (2)  They form a unitary representation of the translation group in two 
dimensional Lorentz space-time,
$$\eqalignno {M^{r\pm}_{\omega - \overline {\omega}}(\kappa (U+U_0))\,&=\,
\int^\infty_{-\infty}\,M^{r\pm}_{\omega - \omega '}(\kappa U)M^{r\pm}
_{\omega ' - \overline {\omega}}(\kappa U_0)d\omega ';&(6.3)\cr}$$
\medskip

\noindent (3)  They are eignfunctions of the Lorentz boosts (see below)
$${\partial \over {\partial \tau}}M^{j\pm}_\omega \,=\,-i\omega M^{j\pm}_\omega
~~~~j\,=\,a, r\,.$$
\medskip

\noindent Furthermore their $K-G$ normalization, Eq. (2.4), is also the same
$$\eqalignno {(M^{j\pm}_\omega ,M^{j\pm}_{\omega '})\,&=\,\pm 2\delta 
(\omega - \omega ')~~~~j\,=\,a,r &(6.4a)\cr}$$
\medskip

\noindent In addition
$$\eqalignno {(M^{a\pm}_\omega ,M^{r\pm}_{\omega '})\,&=\,0&(6.4b)\cr}$$
\medskip

\noindent as one might have expected.
\medskip

The Rindler {\it coordinate representatives} of the $M-M$ mode are readily obtained 
by inserting the expressions from Eqs. (4.1) into the definitions, Eqs. (6.2).  
The resulting four coordinate representatives of the retarded $M-M$ modes are
$$\eqalignno {M^{r\pm}_\omega (\kappa U)\,&=\,{1 \over {2\pi}}\Gamma (i\omega )
(\kappa \xi )^{-i\omega}\,e^{-i\omega \tau}\left \{ 
\matrix {e^{\pm \pi \omega /2}&in&I&and &F\cr
         e^{\mp \pi \omega /2}&in&II&and&P}\right.&(6.5b)\cr}$$
\medskip

Here $\Gamma$ is the gamma function.  Except for a normalization factor 
$\sqrt {2}$ these expressions agree with those already known for $I$ and 
$II^{11}$.
\vfill\eject

\noindent {\bf VII.  CONCLUSION}
\medskip

The global Minkowski Bessel modes are the Poincare group harmonics on the 
Lorentz $2$-plane.  They are analogous to the large $\ell$ limit of the 
$Y^\ell_m(\theta , \phi )$, the rotation group harmonics on the $2$-sphere.  
The three most important ideas of this paper are expressed by {\it Properties} 
1-3.  They hold both for the $M-B$ modes and the $M-M$ modes.  The two sets are 
related by Eqs. (6.2).  The asymptotic approximation of a mode sum as a mode 
integral is exhibited by Eq. (5.22) for a physically interesting  coordinate 
representative.

Finally let us mention three observations which are more 
technical in nature:
\medskip

\noindent (1)  The conventional group theoretic treatment of the MacDonald 
functions $K_{i\omega}$ and the Hankel functions $H_{i\omega}$ view them as a representation 
of translations in the Minkowski plane$^{12}$.  This does not express the true 
state of affairs.  Actually it is the global Minkowski Bessel modes which play 
this role.  The functions $K_{i\omega}$ and $H_{i\omega}$ only refer to the 
{\it coordinate representatives} of these $M-B$ modes in each of the respective Rindler 
charts ("sectors").
\medskip

\noindent (2)  The group composition properties, Eqs. (2.7) and (6.3) for the 
global $M-B$ and $M-M$ modes imply numerous composition theorems for the Bessel 
related functions $(K_{i\omega}, H^1_{i\omega}$ and \hfil\break $H^
2_{i\omega})^{13}$  and the gamma function $\Gamma (i\omega)^{14}$.   These theorems 
are now so easy to obtain 
because one merely has to insert the coordinate representatives, Eqs. (4.2) - 
(4.3) and Eqs. (6.5) into the group composition laws.  
\medskip

\noindent (3)  If one recalls that 
$$\eqalign {B^{\pm}_{\omega - \omega '}(2\kappa U,0)\,&=\,M^{r\pm}_{\omega - 
\omega '}(\kappa U)\cr
\noalign{\hbox{and}}
B^{\pm}_{\omega - \omega '}(0,\kappa V)\,&=\,M^{a \pm}_{\omega - \omega '}
(\kappa V)\,,\cr}$$
\medskip

\noindent then inserting the corresponding coordinate representatives into the 
group composition law Eq. (2.7) yields integral relations$^{15}$ between the 
Bessel related functions and the gamma function.
\vfill\eject

\newcount\bibno\bibno=1
\def\bib#1\par{\hangindent 3em\hangafter 1\noindent\hbox to 3em{\hfil\number
\bibno.\quad}#1\par\smallskip\advance\bibno by 1}

\noindent {\bf REFERENCES}.
\medskip

\bib {U. H. Gerlach, "The Rindler Condensate" in {\it Fourth Marcel Grossman 
Meeting on General Relativity}, R. Ruffini, ed; (Elservier Science Publishers 
B. V. , 1986)  p 1129.}
\smallskip

\bib {U. H. Gerlach, "Photon Condensation", Ohio State report, 1988 (unpublished).}
\smallskip

\bib {S. A. Fulling, Phys. Rev. D $\underline {7}$, 2850 (1973).}
\smallskip

\bib {W. G. Unruh, Phys. Rev. D $\underline {14}$, 870 (1986).}
\smallskip

\bib {W. Israel, Phys. Lett. $\underline {57 A}$, 107 (1976).}
\smallskip

\bib {A. Sommerfeld, {\it Partial Differential Equations in Physics}, (Academic 
Press, New York 1964) p 84.}
\smallskip

\bib {W. Rindler, Am. J. Phys. $\underline {34}$ 1174 (1966).}
\smallskip

\bib {See, for example C. W. Misner, K. S. Thorne, and J. A. Wheeler, {\it 
Gravitation} (Freemann and Co., San Francisco, 1973) Ch. 6.}
\smallskip

\bib {See, for example, Problem 4.8 in I. Stakgold, {\it Boundary Value 
Problems of Mathematical Physics}, Vol. I (Mac-Millan Co., New York, 1967) 
p. 278.} 
\smallskip

\bib {See References 8, p 665.}
\smallskip

\bib {R. Hughes, Ann. Phys. $\underline {162}$, 1 (1985).}
\smallskip

\bib {N. Vilenkin, {\it Special Functions and the Theory of Group Representations}, 
Translations of Mathematical Monographs, Vol. 22, (American Mathem. Soc.; 
Providence, R. I. 1968) p 254-261, p 275-279.}
\smallskip

\bib { Ibid p 275-279}
\smallskip

\bib {Ibid p 241-242}
\smallskip

\bib {See, e.g., idid. p 269, 272-274.}

\end